\documentclass[aip,jcp,reprint]{revtex4-1}
\usepackage{epsfig}
\usepackage{graphicx}
\usepackage{epstopdf}
\usepackage{amsmath}
\usepackage{amssymb}
\usepackage{amsfonts}  
\usepackage{fancyhdr}
\usepackage{bm}
\usepackage{dcolumn}
\usepackage{subfigure}
\newcommand{\intmat}{\mathbf{\mathcal{M}}}

\newcommand{\bra}[1]{\langle #1|}
\newcommand{\ket}[1]{|#1\rangle}
\newcommand{\braket}[2]{\langle #1|#2\rangle}

\newcommand{\half}{\frac{1}{2}}
\newcommand{\rvec}{\mathbf{R}}

\newcommand{\Pvec}{\mathbf{P}}

\newcommand{\xvec}{\mathbf{x}}

\newcommand{\pvec}{\mathbf{p}}

\newcommand{\proj}{\mathcal{P}}

\newcommand{\dbydx}[2]{\frac{\partial #1}{\partial #2}}

\begin{document}
\title{Mapping Variable Ring Polymer Molecular Dynamics: A Path-Integral Based Method for Nonadiabatic Processes}
\author{Nandini Ananth}
\affiliation{Department of Chemistry and Chemical Biology, 
Cornell University, Ithaca, New York, 14853, USA}
%\author{Artur R. Menzeleev}
%\author{Thomas F. Miller III}
%\affiliation{Division of Chemistry and Chemical Engineering, 
%California Institute of Technology, Pasadena, California 91125, USA}
\date{\today}
\begin{abstract}
  We introduce mapping-variable ring polymer 
  molecular dynamics (MV-RPMD), a model dynamics for the 
  direct simulation of multi-electron 
  processes. An extension of the RPMD idea, this method is based on 
  an exact, imaginary time path-integral representation 
  of the quantum Boltzmann operator using continuous 
  Cartesian variables for both electronic states and 
  nuclear degrees of freedom. 
  We demonstrate the accuracy of the MV-RPMD approach
 in calculations of real-time, thermal correlation functions
 for a range of two-state single-mode model systems with 
 different coupling strengths and asymmetries. Further, we show 
 that the ensemble of classical trajectories employed in these 
 simulations preserves the Boltzmann distribution and provides
 a direct probe into real-time coupling between electronic state 
 transitions and nuclear dynamics.
\end{abstract}
\maketitle
\section{Introduction}
Understanding the mechanisms of thermal and photochemical 
charge and energy transfer reactions is a key step towards 
the rational design of energy-efficient devices including
organic photovoltaics, transition metal catalysts for 
water-splitting, and molecular motors. 
\cite{jlb04,mbs10,mm08,pc10,ric98,jld05}
The development of novel theoretical methods to perform 
large-scale simulations of coupled electronic and nuclear 
dynamics in the condensed phase, therefore, remains an important challenge.
%\cite{erb95,fw91,bs91,bs92,am09,tfm08,amw04,hn08,fkf76}.

Exact quantum methods are numerically intractable for 
large systems, necessitating the development of approximate methods 
that exhibit favorable scaling in computational effort with
system size. An additional challenge in the case of 
nonadiabatic charge and energy transfer is the accurate simulation
of coupled electronic and nuclear motion, a challenge best met
by using a consistent dynamic framework for all system 
degrees of freedom. \cite{er98,jcb00,whm09, jc99}
In contrast to mixed quantum-classical approaches,
\cite{pp69,rk99,ad00,jct90,ovp97,awj02,mfh95,yhw07,
pe27,cz04,jct00,ak12,cds99,jll02} semiclassical methods 
\cite{,xs97,na07,whm09,sb052,ph10} 
provide an even-handed treatment 
of coupled nuclear and electronic dynamics but are numerically 
demanding, and their applications, thus far, have been limited to
small systems. 

Ring polymer molecular dynamics 
(RPMD) \cite{cra04,sh13} provides an attractive alternative: 
both electronic and nuclear degrees of freedom are described using a 
position-space path-integral (PI) representation,  
and real-time dynamic information is obtained 
from classical MD trajectories. RPMD has been used 
extensively to characterize nuclear quantum effects
in the condensed phase; \cite{tfm05,tem08,sh09,nb11}
more recent applications to electron transfer and 
proton-coupled electron transfer demonstrate the 
success of RPMD in large-scale, atomistic 
simulations of nonadiabatic processes. \cite{tfm08,arm10,arm11,jsk13}
However, the absence of discrete electronic state variables in the 
RPMD formulation restricts its application to single-electron processes. 
It is clear that new methods are required for atomistic simulations of 
nonadiabatic reactions in photochemistry and, more generally,
for multi-electron chemistry.

In this paper, we present mapping-variable (MV)-RPMD,
 a novel method that preserves the desirable characteristics of 
 RPMD while explicitly including quantized electronic state dynamics.
The first step towards this RPMD-like dynamics is the construction
of an exact, phase-space PI representation for the quantum 
Boltzmann distribution (QBD) of an $N$-level system, 
using continuous variables to represent both electronic and 
nuclear degrees of freedom. We map discrete electronic
state variables to continuous Cartesian variables 
following the Stock-Thoss (ST) protocol.\cite{hdm79,gs97} 
In earlier work, Ananth and Miller \cite{na10} described PI 
discretization of the Boltzmann operator for an $N$-level system 
in the mapping framework
using a projection operator that constrains the electronic position
variables to the mapping subspace. 
Here, we construct an exact, phase-space PI representation for the 
QBD by properly constraining both electronic position and momentum 
variables to the mapping subspace. 
This is accomplished by performing a Wigner transform 
of the trace over electronic variables that includes the projection operator.
Classical equations of motion for the MV-RPMD trajectories are
generated from the resulting expression for the QBD, and these
trajectories are used in the calculation of exact equilibrium 
properties and approximate real-time dynamic correlation functions.

We numerically demonstrate the accuracy of MV-RPMD in calculations of 
thermal correlation functions (TCFs) for a series of two-state systems 
coupled to a single nuclear mode with different coupling strengths and 
asymmetries. We show that our method is consistently better than
a mean-field approximation to RPMD in describing nonadiabatic 
dynamics for systems with weak and intermediate coupling strengths. 
Further, we show that MV-RPMD trajectories can be used as a 
direct probe into real-time changes in electronic
state populations and nuclear positions. 
%\vspace{-0.2in},
\section{Theory}
%\vspace{-0.2in}
\subsection{Position-space PI discretization in the mapping framework}
The Hamiltonian for a general, $N$-level system is written as  
\begin{equation}
	H = \frac{\Pvec^T\cdot\Pvec}{2M}+V_0(\rvec)+\half \sum_{n,m=1}^N
\ket{\psi_n}V_{nm}(\rvec)\bra{\psi_m},
\label{eq:st_ham}
\end{equation}
where $(\rvec,\Pvec)$ represent nuclear positions and momenta, 
$M$ is the nuclear mass, $V_0(\rvec)$ is the electronic 
state-independent potential energy, and $V_{nm}(\rvec)$ 
are elements of the diabatic potential energy matrix.
Following the ST mapping protocol, \cite{gs97} $N$ discrete electronic
states are mapped to $N$ independent, singly excited oscillator (SEO) states,
\begin{eqnarray}
	\ket{\psi_n}\bra{\psi_m}&\rightarrow& a_n^+a_m\\
	\ket{\psi_n}&\rightarrow&
	\ket{0_1,\cdots,1_n,\cdots 0_N},
	\label{eq:stmap}
\end{eqnarray}
where $a_n,\;a_n^+$ are boson creation and annihilation operators 
that obey the commutation rules $[a_n^+,a_m]~=~\delta_{nm}$. 
In the rest of this paper we use the notation 
$\ket{n}~\equiv~\ket{0_1,\cdots,1_n,\cdots 0_N}$ 
to represent the SEO states that
are the product of $N$ independent harmonic oscillators, 
with $(N-1)$ in the ground state and the $n^\text{th}$ oscillator in 
the first excited state.

The canonical partition function for this system is written as
\begin{equation}
Z=\text{Tr}\left[ e^{-\beta H} \right],
	\label{eq:}
\end{equation}
where $\beta=1/kT$.
PI discretization of the Boltzmann operator in the mapping framework 
is performed by inserting $P$ copies of an identity that 
preserves the mapping subspace, \cite{na10}
\begin{equation}
	\mathbf{I}=\int d\rvec \int d\xvec
	\ket{\xvec,\rvec}\bra{\xvec,\rvec}
	\proj,
	\label{eq:identity}
\end{equation}
where $\proj=\sum_{n}\ket{n}\bra{n}$ is the projection 
operator in the SEO basis, and $\xvec$ represents electronic 
position variables.

Using the identity in Eq.~(\ref{eq:identity}), we obtain a PI
expression for the canonical partition function,
\begin{eqnarray}
\nonumber
Z&=&\int d{\{\rvec_\alpha\}}
\int d{\{\xvec_\alpha\}}\\
&& \prod_{\alpha=1}^P \bra{\xvec_\alpha,\rvec_\alpha}
\proj e^{-\beta_P H} \proj \ket{\xvec_{\alpha+1},\rvec_{\alpha+1}},
\label{eq:disc_boltz}
\end{eqnarray}
where $\beta_P=\beta/P$, and we have introduced the notation 
$\int d\{\xvec_\alpha\}=\int d\xvec_1\cdots\int d\xvec_P$.
Using the Trotter approximation, \cite{et58} electronic-state
independent nuclear matrix elements can be evaluated to yield
\begin{eqnarray}
\nonumber
Z&\propto&\lim_{P\to\infty}
\int d{\{\rvec_\alpha\}}
\left(\prod_{\alpha=1}^P \;\mathcal{A}_\alpha \right)
\\
&\times&\int d{\{\xvec_\alpha\}}
\prod_{\alpha=1}^P \;
%\text{Tr}\left[
%\prod_{\alpha=1}^P 
%\ket{\xvec_\alpha}\bra{\xvec_\alpha}
\bra{\xvec_\alpha}
\proj\;e^{-\beta_P \mathbf{V}(\rvec_\alpha)}\proj
%\right]
 \ket{\xvec_{\alpha+1}},
\label{eq:nuc_disc_boltz}
\end{eqnarray}
where 
\begin{equation}
\mathcal{A}_\alpha=
e^{-\beta_P\;V_0(\rvec_\alpha)}
e^{-\frac{MP}{2\beta}
(\rvec_\alpha-\rvec_{\alpha+1})^T\cdot
(\rvec_\alpha-\rvec_{\alpha+1})}\;,
\label{eq:formula_a}
\end{equation}
 $\mathbf{V}(\rvec_\alpha)$ is the diabatic potential 
energy matrix, and we set $\hbar~=~1$ throughout this paper. 
The proportionality sign in 
Eq.~(\ref{eq:nuc_disc_boltz}) indicates that we neglect 
pre-multiplicative constants.
Electronic matrix elements are evaluated using the SEO wavefunction,
\begin{equation}
\braket{\xvec}{n}=\frac{\sqrt{2}}{\pi^{N/4}}\; \left[ \xvec\right]_n
e^{-\half \xvec^T\cdot\xvec},
\label{eq:seo_wvfn}
\end{equation}
where $\left[\cdot\right]_n$ indicates the $n^\text{th}$ component
of the enclosed vector,
and the Boltzmann matrix elements in SEO states are 
evaluated using a high-temperature approximation,\cite{dc87}
\begin{equation}
	\bra{n}	
	e^{-\beta_P\mathbf{V}(\rvec)}\ket{m}=\intmat_{nm}(\rvec),
	\label{eq:boltz_mat}
\end{equation}
where
\begin{equation}
	\intmat_{nm}(\rvec)=\left\{
	   \begin{array}{cc}
	   e^{-\beta_P V_{nn}(\rvec)}&,n=m. \\
	   -\beta_P V_{nm}(\rvec)\;e^{-\beta_P V_{nn}(\rvec)}
	   &,n\ne m.\\
	   \end{array}\right.
	\label{eq:intmat}
\end{equation}
The resulting expression is the previously derived PI-ST representation
\cite{na10} for the canonical partition function,
\begin{eqnarray}
	\nonumber
	Z &\propto& 
	\lim_{P\to\infty}
	%\left(\frac{2MP}{\hbar^2\beta\pi^{N+1}}\right)^\frac{fP}{2}
	\int d\left\{\rvec_\alpha \right\} 
	\int d\left\{\xvec_\alpha\right\}\\
	&\times&
	\prod_{\alpha=1}^P 
	\mathcal{A}_\alpha\;\mathcal{F}_\alpha\;\mathcal{G}_\alpha,
	\label{eq:pist}
\end{eqnarray}
where
\begin{eqnarray}
&&\mathcal{F}_\alpha
=\xvec_\alpha^T\;
\intmat(\rvec_\alpha)
\;\xvec_{\alpha+1},
\label{eq:formula_f}
\\
&&\mathcal{G}_\alpha=
e^{-\xvec_\alpha^T\cdot\xvec_\alpha},
\label{eq:formula_g}
\end{eqnarray}
and $\mathcal{A}_\alpha$ is defined in Eq.~({\ref{eq:formula_a}).

Thus far, we have reviewed the PI discretization approach used 
to derive the PI-ST representation; \cite{na10} going forward, we start
with Eq.~(\ref{eq:nuc_disc_boltz}) and construct a phase-space
PI expression for the QBD by introducing 
momentum-space integrals in both the nuclear and 
electronic variables.

\subsection{Phase-space PI representation using mapping variables}
We introduce nuclear momentum variables using 
normalized Gaussian integrals\cite{mp84}
\begin{equation}
	I_N=\left(
\frac{2\pi M'}{\beta_P}\right)^{\frac{fP}{2}}
\int d\{\Pvec_\alpha\} e^{-\frac{\beta_P}{2M'}\sum_{\alpha=1}^P
\Pvec_\alpha^T.\Pvec_\alpha},
	\label{eq:nuc_mom}
\end{equation}
where $f$ is the number of nuclear degrees of freedom.
In keeping with the RPMD formalism, \cite{cra04} the fictitious mass term 
in Eq.~(\ref{eq:nuc_mom}) is chosen to be the 
physical mass of the nuclei, $M'=M$. 
Unfortunately, introducing electronic momentum integrals is not as 
straightforward -- both electronic position and momentum variables must
be constrained simultaneously to the mapping subspace.
We achieve this by replacing the trace over electronic path-variables
with the corresponding Wigner transforms.

Consider the integral over electronic path-variables
in Eq.~(\ref{eq:nuc_disc_boltz}),
\begin{eqnarray}
\label{eq:elec_int}
	I_E &=&
	\int d\xvec_1\cdots\int d\xvec_P\;
	\bra{\xvec_1}\proj
	e^{-\beta_P\mathbf{V}(\rvec_1)}\proj
	\ket{\xvec_2}\\
	\nonumber
	&\times&
	\bra{\xvec_2}\proj
	e^{-\beta_P\mathbf{V}(\rvec_2)}\proj
	\ket{\xvec_3}
	\cdots
	\bra{\xvec_P}\proj
	e^{-\beta_P\mathbf{V}(\rvec_P)}\proj
	\ket{\xvec_1},
	\end{eqnarray}
where the integral over $\xvec_1$ can be replaced by a
trace,
\begin{eqnarray}
I_E&=&
\int d\xvec_2\cdots\int d\xvec_P\;
\text{Tr}\left[ \hat{S}\right]_1.
\label{eq:trace_rep}
\end{eqnarray}
The composite operator $\hat{S}$ in Eq.~(\ref{eq:trace_rep}) is 
introduced to represent a product of operators,
\begin{eqnarray}
\label{eq:as_trace}
\hat{S}&=&
\proj
e^{-\beta_P\mathbf{V}(\rvec_1)}\proj
\ket{\xvec_2}\\
\nonumber
&\times&
\bra{\xvec_2}\proj
e^{-\beta_P\mathbf{V}(\rvec_2)}\proj
\ket{\xvec_3}
\cdots
\bra{\xvec_P}\proj
e^{-\beta_P\mathbf{V}(\rvec_P)}\proj,
\end{eqnarray}
and we use the notation $\text{Tr}\left[ \cdot \right]_\alpha$ 
to indicate a trace over the $\alpha^\text{th}$ electronic path-variable. 

In the phase-space formulation of quantum mechanics, \cite{hw27,ew32a,ew32b,jem49} 
the trace over an operator $\hat{O}$ is expressed as a phase-space 
integral of the corresponding Wigner function,
\begin{equation}
	\text{Tr}[\hat{O}]=\frac{1}{(2\pi)^N}
	\int d\xvec\int d\pvec\;O(\xvec,\pvec),
	\label{eq:def_wt}
\end{equation}
where the Wigner function is obtained from the expression
\begin{eqnarray}
	O(\xvec,\pvec)&=&\int d \Delta\xvec
	\bra{\xvec-\frac{\Delta\xvec}{2}}\hat{O}
	\ket{\xvec+\frac{\Delta\xvec}{2}}
	e^{i\pvec^T\cdot\Delta\xvec}.
	\label{eq:def_wd}
\end{eqnarray}

Using Eq.~(\ref{eq:def_wt}) and Eq.~(\ref{eq:def_wd}) the 
trace in Eq.~(\ref{eq:trace_rep}) can be written as a phase-space 
integral of the form 
\begin{eqnarray}
\nonumber
\text{Tr}\left[ \hat{S} \right]_1 &=&
\frac{1}{(2\pi)^N}\int d\xvec_1\int d\pvec_1
\int d\Delta\xvec_1 \\
&& \;\;
\bra{\xvec_1-\frac{\Delta\xvec_1}{2}}
\hat{S}\ket{\xvec_1+\frac{\Delta\xvec_1}{2}}\;e^{i\pvec_1^T\cdot\Delta\xvec_1}.
\label{eq:wt_s}
\end{eqnarray}
Substituting Eq.~(\ref{eq:wt_s}) back into Eq.~(\ref{eq:trace_rep}), we 
obtain
\begin{eqnarray}
\nonumber
I_E&=&
\frac{1}{(2\pi)^N}
\int d\xvec_1\int d\pvec_1
\int d\Delta\xvec_1
\int d\xvec_2\cdots\int d\xvec_P\\
&&
\bra{\xvec_1-\frac{\Delta\xvec_1}{2}}
\hat{S}\ket{\xvec_1+\frac{\Delta\xvec_1}{2}}
\;e^{i\pvec_1^T\cdot\Delta\xvec_1}.
\label{eq:ps}
\end{eqnarray}
We use the definition of operator $\hat{S}$ in Eq~.(\ref{eq:as_trace})
to rewrite the integral over $\xvec_2$ as a trace,
replace the trace by a phase-space integral over the 
corresponding Wigner distribution, and repeat this procedure
until all $P$ position-space integrals have been replaced by phase-space
integrals and $P$ additional $\{\Delta\xvec_\alpha\}$ integrals,
\begin{eqnarray}
\label{eq:psp}
&&I_E=
\frac{1}{(2\pi)^{PN}}
\int d\{\xvec_\alpha\}
\int d\{\pvec_\alpha\}
\int d\{\Delta\xvec_\alpha\}\\
&&\;\;
\nonumber
\prod_{\alpha=1}^P
\bra{\xvec_\alpha-\frac{\Delta\xvec_\alpha}{2}}
\proj e^{\beta_P \mathbf{V}(\rvec_\alpha)}\proj
\ket{\xvec_{\alpha+1}+\frac{\Delta\xvec_{\alpha+1}}{2}}
\;e^{i\pvec_\alpha^T\cdot\Delta\xvec_\alpha}.
\end{eqnarray}
We analytically integrate over the variables 
$\{\Delta\xvec_\alpha\}$, described in detail
in the Appendix, to reduce Eq.~(\ref{eq:psp}) to 
an integral over electronic phase-space variables only,
\begin{eqnarray}
\nonumber
I_E&=&
\frac{1}{(2\pi)^{PN}}
\int d\{\xvec_\alpha\}
\int d\{\pvec_\alpha\}\\
&\times&
\text{Tr}\left[\;{\bm \Gamma}
\right]\;
e^{-\sum_{\alpha=1}^P
\left( 
\xvec_\alpha^T\cdot\xvec_\alpha
+\pvec_\alpha^T\cdot\pvec_\alpha
\right)},
\label{eq:psint_ie}
\end{eqnarray}
where
\begin{eqnarray}
	{\bm \Gamma}=\prod_{\alpha=1}^P\left(
	{\bf{\mathcal{C}}}_\alpha-\half\;{\bf \mathcal{I}}\right)
	\intmat(\rvec_\alpha).
	\label{eq:gam_def}
\end{eqnarray}
In Eq.~(\ref{eq:gam_def}), the complex matrix
\begin{eqnarray}
\label{eq:formula_cmat}
\bf{\mathcal{C}}_\alpha &=& 
\left( \xvec_\alpha+i\pvec_\alpha \right)
\otimes
\left( \xvec_\alpha-i\pvec_\alpha \right)^T,
\end{eqnarray}
$\mathcal{I}$ is the identity matrix, 
and $\intmat(\rvec)$ is previously defined in Eq.~(\ref{eq:intmat}).

Replacing the electronic integral in Eq.~(\ref{eq:nuc_disc_boltz}) 
with Eq.~(\ref{eq:psint_ie})
and introducing the nuclear momentum variables with
Eq.~(\ref{eq:nuc_mom}), we 
obtain an exact, phase-space PI 
representation for the QBD of an $N$-level system,
\begin{eqnarray}
	\nonumber
	Z &\propto& \lim_{P\to\infty}
	%\left(\frac{2MP}{\beta\pi^{N+1}}\right)^\frac{fP}{2}
	\int d\left\{\rvec_\alpha \right\} 
	\int d\left\{\Pvec_\alpha \right\}
	\int d\left\{\xvec_\alpha\right\}
	\int d\left\{\pvec_\alpha\right\}\\
	&\times&
	e^{-\beta_P H_P
	\left( \{\rvec_\alpha\},\{\Pvec_\alpha\},
	\{\xvec_\alpha\},\{\pvec_\alpha\} \right)
	}
	\text{sgn}\left( \Theta \right).
	\label{eq:final_qbd}
	%\;\text{sgn}\left[ \mathcal{\bar F} \right],
\end{eqnarray}
In Eq.~(\ref{eq:final_qbd}), $H_P$ is the MV-RPMD
Hamiltonian,
\begin{eqnarray}
	H_P = \sum_{\alpha=1}^P \left(
	\mathcal{\bar A}_\alpha\; +
 	\frac{P}{\beta}
	\mathcal{\bar G}_\alpha\right)
	-\frac{P}{\beta}\text{ln}{\bm |}
	\Theta{\bm |},
	%-\frac{iP}{\beta}\Theta.
	\label{eq:mvr_ham}
\end{eqnarray}
where
%In Eq.~(\ref{eq:mv_ham}), we define
\begin{eqnarray}
	\nonumber
	\mathcal{\bar A}_\alpha=
	\frac{MP^2}{2\beta^2}&&
	(\rvec_\alpha-\rvec_{\alpha+1})^T\cdot
	(\rvec_\alpha-\rvec_{\alpha+1})\\
	\label{eq:formula_abar}
	&&\;\;\;\;\;\; 
	+\; V_0(\rvec_\alpha)+\frac{1}{2M}\Pvec_\alpha^T\cdot\Pvec_\alpha
\end{eqnarray}
and 
\begin{eqnarray}
	\label{eq:formula_g}
	\mathcal{\bar G}_\alpha&=&
	\xvec_\alpha^T\cdot\xvec_\alpha+
	\pvec_\alpha^T\cdot\pvec_\alpha.
\end{eqnarray}
Recognizing that the canonical partition 
function is real-valued, the function $\Theta$ in 
Eq.~(\ref{eq:final_qbd}) includes only
the real part of the complex pre-exponential function 
in Eq.~(\ref{eq:psint_ie}),
\begin{eqnarray}
	\Theta&=&
	%\text{Re}\left(\fbar\right)\\
	%&=&
	\text{Re}\left(\text{Tr}\left[\;{\bm \Gamma}\;
	\right]\right).
	\label{eq:int_term_mvr}
\end{eqnarray}

\vspace{0.1in}

\subsection{MV-RPMD trajectories and correlation functions}
The effective MV-RPMD Hamiltonian in 
Eq.~(\ref{eq:mvr_ham}) is used to generate classical, real-space
trajectories that preserve the QBD for an $N$-level system.
The equations of motion for the nuclear and electronic 
variables are
\begin{eqnarray}
\label{eq:mvr_eomi}
\dot{\rvec}_\alpha&=& \frac{\Pvec_\alpha}{M},\\
\nonumber
\dot{\Pvec}_\alpha&=& -\frac{MP}{\beta^2}\left( 2\rvec_{\alpha}-
\rvec_{\alpha+1}-\rvec_{\alpha-1}\right)\\
&&\;\;
-\left( \dbydx{V_0}{\rvec_{\alpha}} \right)
+\frac{P}{\beta\Theta}\left(\dbydx{\Theta}{\rvec_\alpha}\right),\\
\dot{\left[\xvec_{\alpha}\right]}_j&=&\frac{2P}{\beta} \left[\pvec_{\alpha}\right]_j
-\frac{P}{\beta\Theta}\left(
\dbydx{\Theta}{ \left[\pvec_{\alpha}\right]_j}\right),\\
\dot{\left[\pvec_{\alpha}\right]_j}&=&-\frac{2P}{\beta} \left[\xvec_{\alpha}\right]_j
+\frac{P}{\beta\Theta}\left(
\dbydx{\Theta}{ \left[\xvec_{\alpha}\right]_j }\right),
\label{eq:mvr_eomf}
\end{eqnarray}
where, as before, $\left[\cdot\right]_j$ refers to the 
$j^\text{th}$ component of the enclosed electronic variable.

Real-time TCFs in the RPMD framework are identical at 
time zero to the corresponding quantum mechanical Kubo-transformed 
correlation functions, \cite{cra04} as are MV-RPMD TCFs.
Consider the Kubo-transformed nuclear position-position TCF,
\begin{eqnarray}
\label{eq:kt_tcf}
\nonumber
C_{RR}^\text{KT}(t)&=&\frac{1}{\beta Z}
\int_{0}^\beta d\lambda\\
&&\times \;\text{Tr}\left[e^{-(\beta-\lambda) H}
\hat{R}e^{-\lambda H} e^{iHt}\hat{R}e^{-iHt}\right].
\end{eqnarray}
The corresponding MV-RPMD correlation function is 
written as  
\begin{eqnarray}
\label{eq:mvr_tcf}
C_{RR}^\text{MVR}(t)&=&\frac{1}{Z} 
\int d\{\xvec_\alpha\}\int d\{\pvec_\alpha\}
\int d\{\rvec_\alpha\}\int d\{\Pvec_\alpha\}\;\;\;\;\;\\
\nonumber
&\times& 
e^{-\beta_P H_P(\{\xvec_\alpha\},\{\pvec_\alpha\},
\{\rvec_\alpha\},\{\Pvec_\alpha\})}
\bar{\rvec}(0)\bar{\rvec}(t)\text{sgn}\left( \Theta \right)
\end{eqnarray}
where $\bar{\rvec}~=~\frac{1}{P}\sum_{\alpha=1}^P\rvec_\alpha$
is the nuclear center-of-mass coordinate. In our simulations,
 the initial distribution is obtained from standard path-integral
Monte Carlo (PIMC) importance sampling using the function 
$W=e^{-\beta_PH_P}$.
We can write the expression for the real-time TCF in 
Eq.(\ref{eq:mvr_tcf}) as 
\begin{equation}
C_{RR}^\text{MVR}(t)=
\frac
{\left\langle \bar{\rvec}(0)\bar{\rvec}(t)\text{sgn}(\Theta)\right\rangle_W}
{\left\langle\text{sgn}(\Theta)\right\rangle_W},
	\label{eq:mvr_tcf_is}
\end{equation}
where the angular brackets indicate the ensemble average is obtained 
with respect to the function $W$. The nuclear center-of-mass
coordinate, $\bar{R}(t)$, is time-evolved using the equations 
of motion provided in Eqs.~(\ref{eq:mvr_eomi})-(\ref{eq:mvr_eomf}). 
The function $\text{sgn}\left( \Theta \right)$ 
that appears in both the numerator and denominator 
of Eq.~(\ref{eq:mvr_tcf_is}) is constant along a given trajectory.

Instantaneous values of the nuclear center-of-mass and electronic
state populations along a single MV-RPMD trajectory provide insight 
into their relative timescales of motion. 
The average electronic state population in the MV-RPMD framework is 
obtained from
\begin{eqnarray}
\nonumber
\left\langle \proj_n\right\rangle
&=& \frac{1}{Z}
\int d\left\{\rvec_\alpha \right\} 
\int d\left\{\Pvec_\alpha \right\}
\int d\left\{\xvec_\alpha\right\}
\int d\left\{\pvec_\alpha\right\}\;\\
\nonumber
&\times&
e^{-\beta_P H_P
\left( \{\rvec_\alpha\},\{\Pvec_\alpha\}\right)
\left( \{\xvec_\alpha\},\{\pvec_\alpha\}\right)
} \text{sgn}(\Theta)\\
&&\times
\left(
\frac{ \bm \Gamma_{nn}}{\bm \Gamma}
\right),
\label{eq:eq_est}
\end{eqnarray}
where the $\bm \Gamma$ matrix is defined in Eq.~(\ref{eq:gam_def}),
and the ratio on the last line is used to calculate
instantaneous state populations.

\subsection{Mean-field RPMD}
The mean-field (MF) approximation to RPMD is derived 
by integrating over the electronic variables to obtain an effective potential
energy surface for nuclear dynamics. This approximate treatment
of nonadiabatic dynamics is increasingly valid as we move towards 
the strong coupling or adiabatic-limit.
Here we present a derivation starting with the PI-ST expression
for the QBD in Eq.~(\ref{eq:pist}); however, it is possible to obtain
an identical formulation using discrete electronic state variables.
\cite{dem_old}

The integral over electronic matrix elements in Eq.~(\ref{eq:pist}) 
can be evaluated exactly, yielding
\begin{eqnarray}
Z\propto\int d\{\rvec_\alpha\} 
\left( \prod_{\alpha=1}^P\;\mathcal{A}_\alpha\right)
\Theta'\left( \{\rvec_\alpha\} \right)
\label{eq:mf_qbd}
\end{eqnarray}
where $\mathcal{A}_\alpha$ is defined in Eq.~(\ref{eq:formula_a}),
$\intmat(\rvec)$ is defined in Eq.~(\ref{eq:intmat}), and 
\begin{equation}
	\Theta'\left( \{\rvec_\alpha\}\right)=
\text{Tr}\left[ \prod_{\alpha=1}^P \intmat\left( \rvec_\alpha \right) \right].
\label{eq:def_rprime}
\end{equation}
As before, nuclear momenta are inserted using normalized Gaussian integrals
resulting in an exact, phase-space PI representation of the QBD, 
\begin{eqnarray}
\nonumber
Z&\propto&\int d\{\rvec_\alpha\}\int d\{\Pvec_\alpha\} \\
&\times& e^{-\beta_P H_P^\text{MF}
\left( \{\rvec_\alpha\},\{\Pvec_\alpha\}\right)}
\;\text{sgn}\left( \Theta' \right),
\end{eqnarray}
where
\begin{equation}
	H_P^\text{MF}=\bar{\mathcal{A}}_\alpha -
\frac{P}{\beta}\text{ln}\left| \Theta' \right|,
\label{eq:mf_ham}
\end{equation}
and $\bar{\mathcal{A}}_\alpha$ is defined in Eq.~(\ref{eq:formula_abar}).
We note that the function $\Theta'$ in Eq.~(\ref{eq:def_rprime}) 
is always positive for a two-state system. However, 
for a general $N$-level system, $\left( N>2 \right)$, 
it is positive only if the off-diagonal coupling terms in 
the diabatic potential matrix are all positive.

Nuclear dynamics in the MF-RPMD method are generated by 
the Hamiltonian in Eq.~(\ref{eq:mf_ham}),
\begin{eqnarray}
\dot{\rvec}_\alpha&=& \frac{\Pvec_\alpha}{M},\\
\nonumber
\dot{\Pvec}_\alpha&=& -\frac{MP}{\beta^2}\left( 2\rvec_{\alpha}-
\rvec_{\alpha+1}-\rvec_{\alpha-1}\right)\\
&&\;\;
-\left( \dbydx{V_0}{\rvec_{\alpha}} \right)
+\frac{P}{\beta\Theta'}\left(\dbydx{\Theta'}{\rvec_\alpha}\right),
\end{eqnarray}
and the corresponding real-time TCF is written as
\begin{eqnarray}
\label{eq:mf_tcf}
C_{RR}^\text{MF}(t)&=&\frac{1}{Z} 
\int d\{\rvec_\alpha\}\int d\{\Pvec_\alpha\}\\
\nonumber
&\times& 
e^{-\beta_P H_P^\text{MF}(\{\rvec_\alpha\},\{\Pvec_\alpha\})}
\bar{\rvec}(0)\bar{\rvec}(t)\text{sgn}\left( \Theta'\right).
\end{eqnarray}

\section{Results and Discussion}
We calculate real-time, nuclear position-position TCFs
for a two-level system coupled to a single nuclear mode.
This benchmark system was used previously to characterize
semiclassical dynamics initialized to an exact QBD. \cite{na10}
The Hamiltonian for our series of models is
\begin{equation}
	H=\frac{P^2}{2M}+V_0(R)+{\bf{V}}(R),
\end{equation}
%\cite{arm11} is the model system we use 
%to test our MV-RPMD 
where the state-independent potential is $V_0(R)=\half k R^2$.
The diagonal elements of the diabatic potential matrix, ${\bf{V}}(R)$, are
\begin{eqnarray}
\nonumber
V_{11}(R)&=& aR + c,\\
V_{22}(R)&=& -aR,
\end{eqnarray}
and the off-diagonal elements $V_{12}=V_{21}=\Delta$.

We construct a series of four models, three of which are symmetric
and differ only in the strength of the nonadiabatic coupling.
The fourth model is an asymmetric system where one diabatic state
is significantly higher in energy than the other. 
All simulations are performed at a temperature $\beta=1$ a.u., with 
nuclear mass $M=1$ a.u. and potential parameters $k=a=1$ a.u.
The coupling strength and asymmetry for each model are 
provided in Table~\ref{table:params}.
The potential energy
curves for the symmetric and asymmetric models are shown in Fig.~\ref{fig1:potential}.

\begin{figure}[!htb]
   \includegraphics[angle=0,scale=0.3]{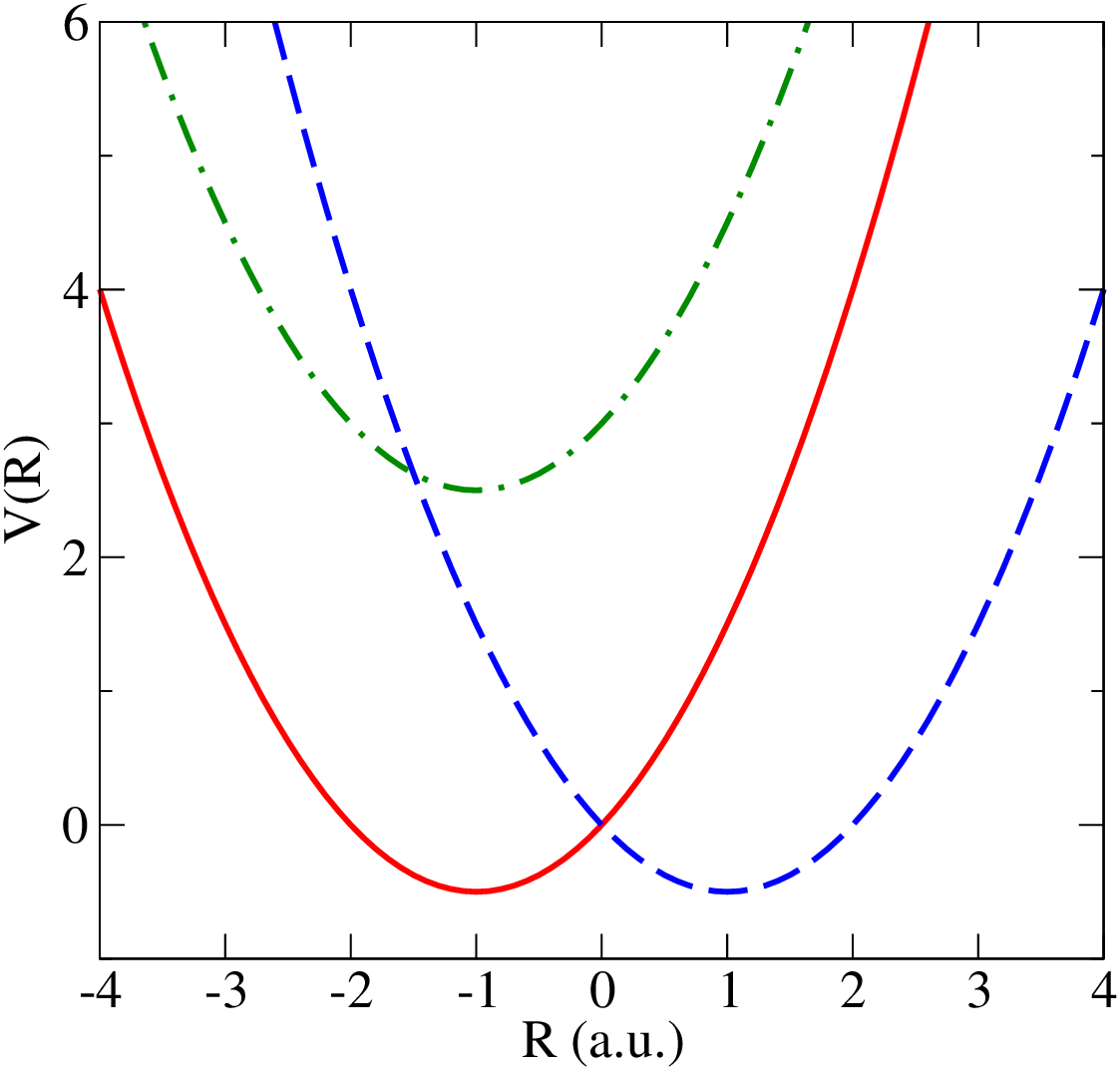}
   \caption{The diagonal elements of the potential energy matrix, 
   $V_{nn}(R)+V_0(R)$, are shown in red (solid line) for electronic state 1
  and in blue (dashed line) for electronic state 2. The green curve 
  (dot-dashed line) corresponds to electronic state 1 for the asymmetric 
  model.}
  \label{fig1:potential}
  \vspace{0cm}
\end{figure}

\begin{center}
\vspace{-0.1in}
\begin{table}[h]
\begin{tabular}{| c | c | c |}
	\hline
	Model & $c$ & $\Delta$ \\ 
	\hline
	I & 0.0 & 0.10 \\ 
	II & 0.0 & 1.00 \\
	III & 0.0 & 10.0 \\
	IV & 3.0 & 0.10 \\
	\hline
\end{tabular}
\caption{Parameters for Models I-IV. All parameters
are specified in atomic units.}
\label{table:params}
\end{table}
%\vspace{-0.5in}
\end{center}
\vspace{-0.2in}

We calculate the nuclear position TCFs for all four model systems.
The MV-RPMD results are obtained by evaluating Eq.~(\ref{eq:mvr_tcf_is}). 
The initial distribution in electronic and nuclear positions and momenta 
for all simulations is sampled using PIMC, 
and a total of $10^5$ points are generated for each model system.
MV-RPMD trajectories are initialized from this initial distribution 
and the equations of motion are integrated for $20$ a.u. using a 
$4^\text{th}$ order Adams-Bashforth-Moulton predictor-corrector integrator.
The initial distribution in the MF-RPMD simulations are sampled 
using PIMC, and a total of $10^3$ points are generated for each model system. 
Trajectories initialized from
this distribution are time evolved for $20$ a.u., and the TCFs are calculated 
using Eq.~(\ref{eq:mf_tcf}). 
We report convergence parameters for both MV-RPMD and MF-RPMD simulations 
in Table~\ref{table:convergence} for all four model systems.
The Kubo-transformed TCF in 
Eq.~(\ref{eq:kt_tcf}) for each model is obtained from a 
numerically exact discrete-variable representation (DVR) grid calculation. \cite{dtc92}

Model I lies in the weak-coupling regime where $\Delta~<<~kT$. This is 
the most physically relevant regime for nonadiabatic electron transfer, 
proton-coupled electron transfer and exciton dynamics, and it is also where
the mean-field approximation breaks down.
In Fig.~\ref{fig:crr_dpt1}, we compare the TCFs obtained from 
MV-RPMD and MF-RPMD simulations with the exact quantum result.
While both simulations are exact at time zero, the MF approximation 
does not hold in this regime, and even at very short times the 
TCF is dramatically different from the exact result.
In contrast, MV-RPMD performs remarkably well: 
the TCF is identical to the quantum result at short times and 
starts to deviate only at longer times.
%The calculations were fully 
%converged with $P=4$ beads and 5000 trajectories.

\begin{figure}[h]
   \includegraphics[angle=0,scale=0.3]{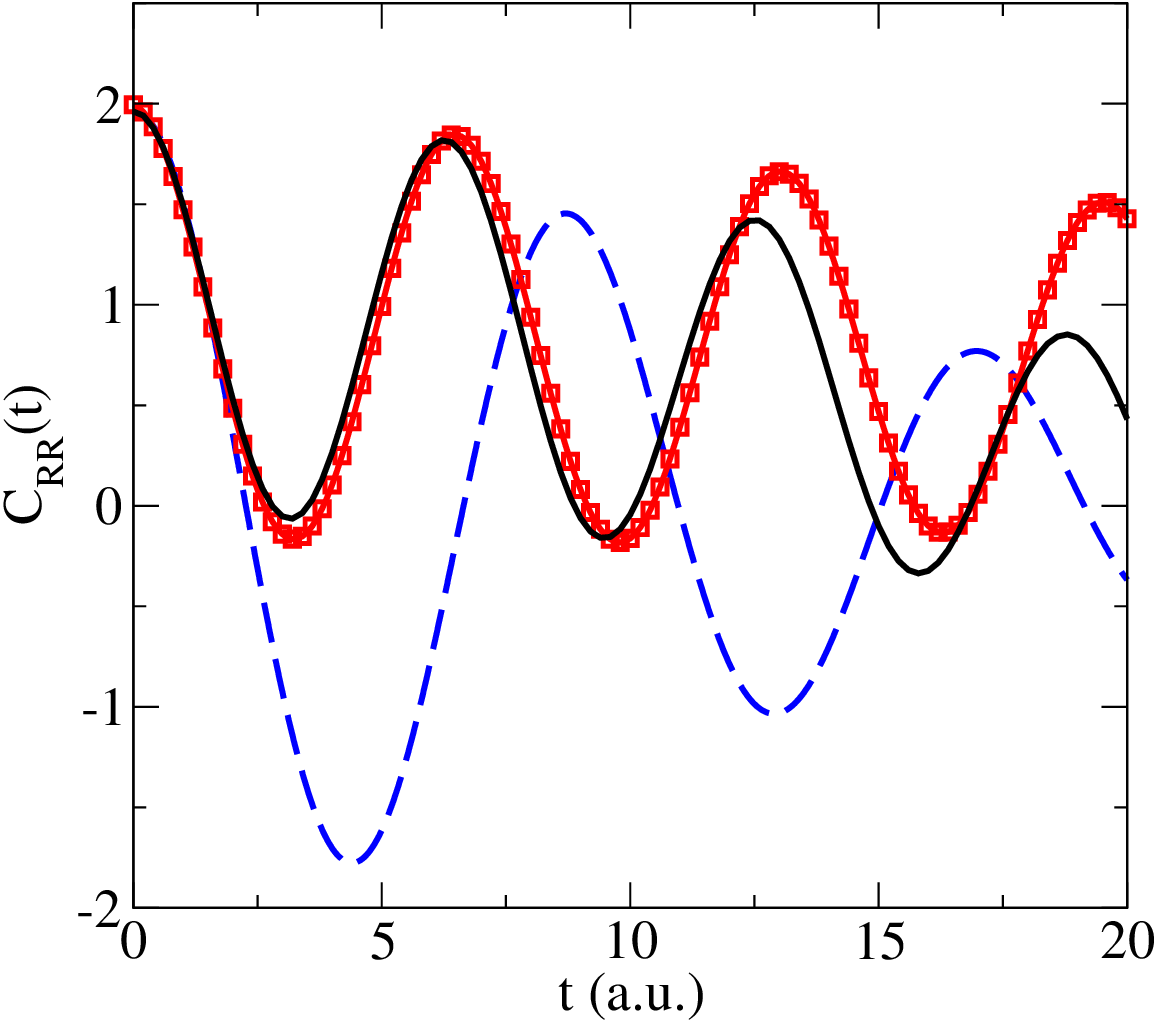}
  \caption{The nuclear position-position TCF for Model I obtained 
  using MV-RPMD is shown in red (solid line with squares), the 
  TCF using MF-RPMD is shown in blue (dashed line),
  and the exact DVR grid calculation is shown in black (solid line).}
  \label{fig:crr_dpt1}
  \vspace{0cm}
\end{figure}

Model II describes a symmetric two-level system with intermediate coupling
strength, $\Delta~\approx~kT$. The nuclear position TCFs for this model
are shown in Fig.~\ref{fig:crr_d1}. The MF-RPMD result is much
improved for this model and correctly captures the timescales of oscillation
in the nuclear positions. MV-RPMD outperforms MF-RPMD again in this regime, 
being nearly identical at short times and deviating slightly at longer times 
from the exact result. 
%In this case, the simulation was converged using $P=5$ beads and 8000 trajectories.

\begin{figure}[h]
   \includegraphics[angle=0,scale=0.3]{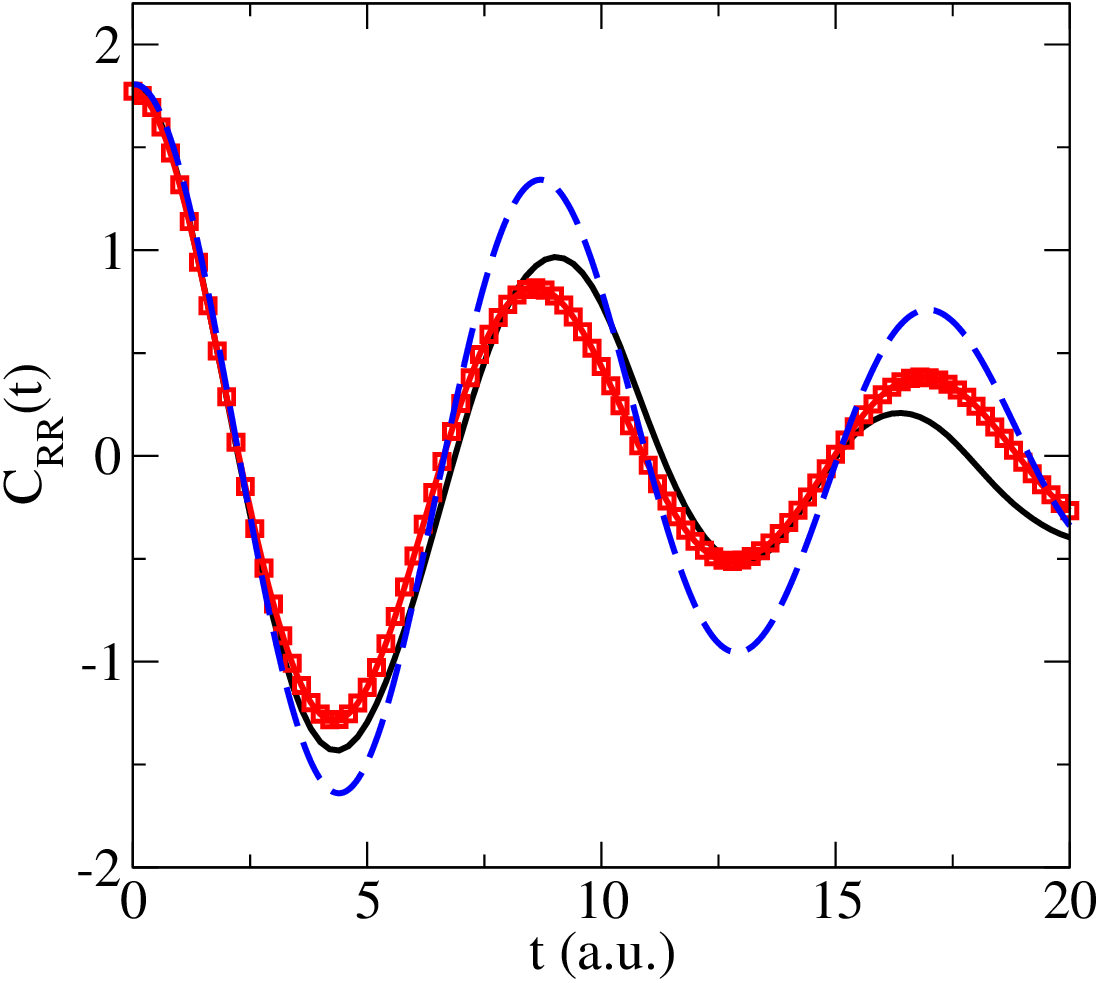}
  \caption{The nuclear position-position TCF for Model II obtained 
  using MV-RPMD is shown in red (solid line with squares), the 
  TCF using MF-RPMD is shown in blue (dashed line),
  and the exact DVR grid calculation is shown in black (solid line).}
  \label{fig:crr_d1}
  \vspace{0cm}
\end{figure}

To confirm that the MV-RPMD trajectories preserve the Boltzmann 
distribution, we calculate averages over the ensemble 
of trajectories used in the $P=5$ bead simulation for Model II. 
The average nuclear center-of-mass coordinate and electronic 
state populations are found to stay constant as a function of time
for the length of our simulation, as expected. 
We demonstrate the potential utility of MV-RPMD for direct dynamics 
by calculating instantaneous values of electronic state populations 
from Eq.~(\ref{eq:eq_est}) and 
nuclear center-of-mass coordinate along a single, representative trajectory.
In Fig.~\ref{fig:single_traj}  we show that for this intermediate 
coupling regime electronic state populations oscillate on the timescale 
of the nuclear vibrational motion.

\begin{figure}[h]
   \includegraphics[angle=0,scale=0.3]{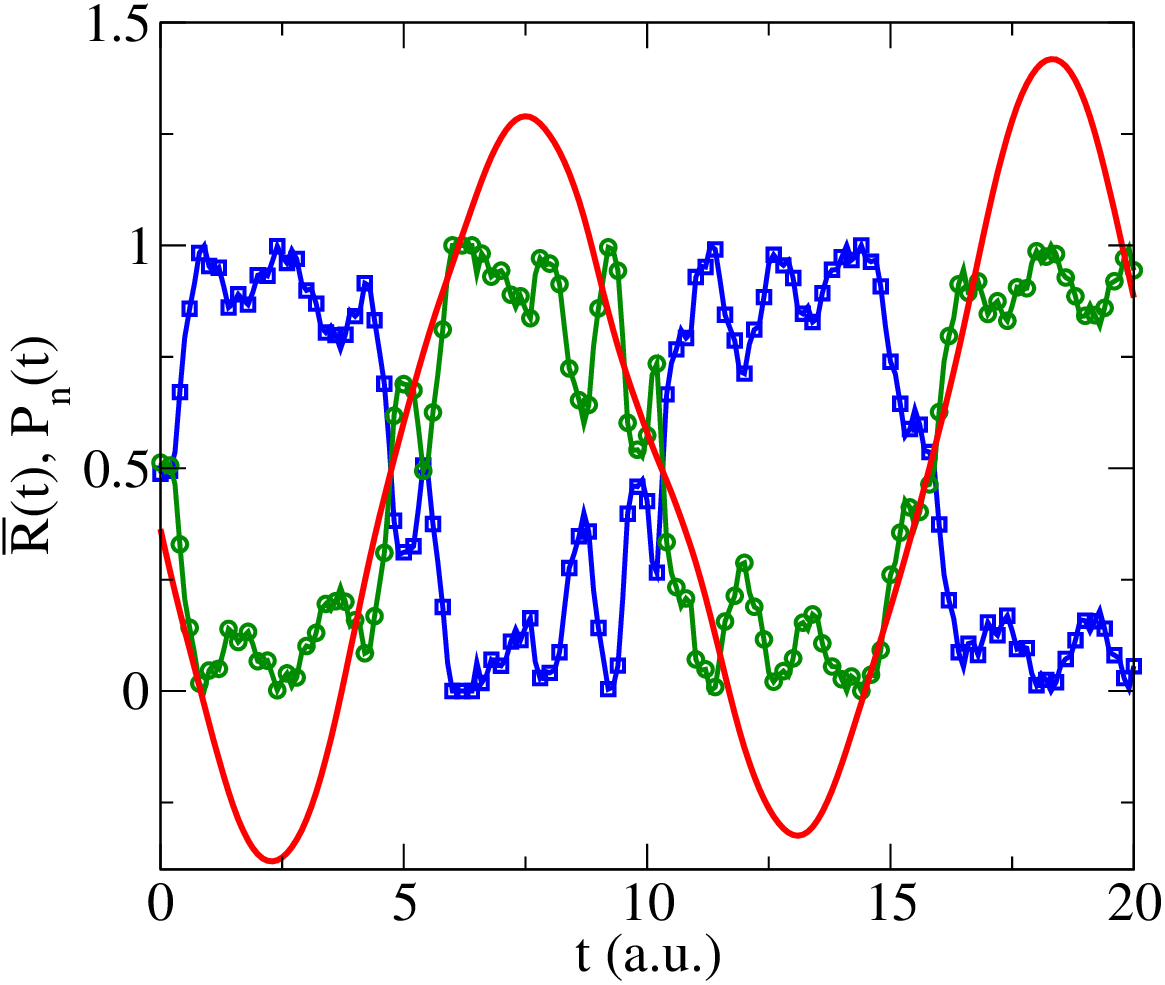}
  \caption{The instantaneous values of electronic state populations
  for Model II are shown in blue (with squares) for state 1 and in 
  green (with circles) for state 2, and the nuclear center-of-mass
coordinate is shown in red (solid line) along a single MV-RPMD 
trajectory.
The nuclear coordinate is scaled and shifted such that the diabatic
potentials appear to cross at $R=0.5$ rather than $R=0$ for clarity.}
  \label{fig:single_traj}
  \vspace{0cm}
\end{figure}

Model III represents the strong coupling regime where $\Delta~>>~kT$.
The mean-field approximation works well in this regime; the results
from our simulation were identical to the exact quantum result and 
are not shown here. 
In Fig.~\ref{fig:crr_d10}, we compare the MV-RPMD correlation function
with the exact quantum results, and find that it is nearly identical  
at all times. 
%The calculations, were however, harder to 
%converge and required $P=10$ path beads and a total of $20000$ trajectories. 

\begin{figure}[!htb]
   \includegraphics[angle=0,scale=0.3]{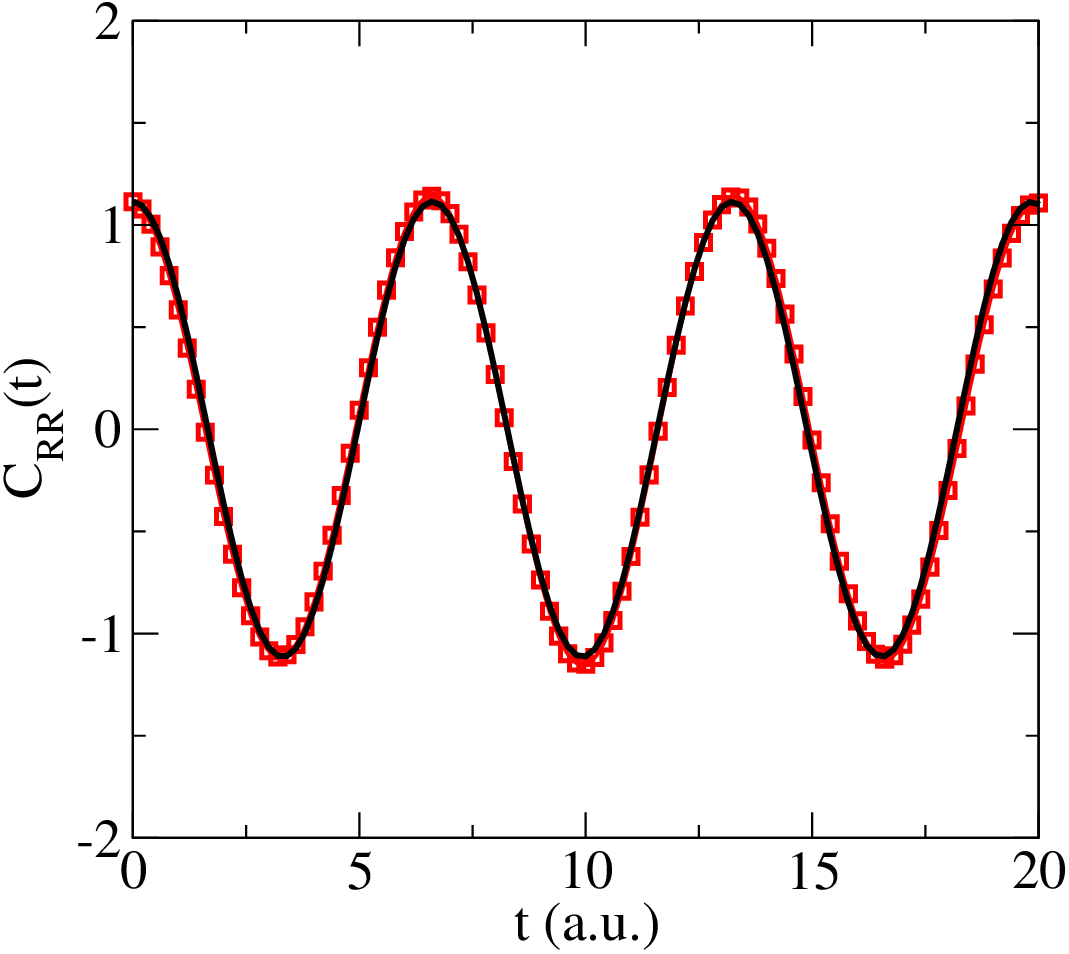}
  \caption{The nuclear position-position TCF for Model III obtained 
  using MV-RPMD is shown in red (solid line with squares)
  and the exact DVR grid calculation is shown in black (solid line).
The MF-RPMD TCF is not plotted, as it is also nearly identical with 
the exact quantum TCF.}
  \label{fig:crr_d10}
  \vspace{0cm}
\end{figure}

\begin{center}
\vspace{-0.1in}
\begin{table}[b]
\begin{tabular}{ |c|c|c|c|c |}
	\hline
	Model    & \multicolumn{2}{c|}   {MV-RPMD} & \multicolumn{2}{c|}{MF-RPMD}\\
	%\cline{2-5}
	  & $P$ & $N_\text{T}$ & $P$ & $N_\text{T}$ \\ 
	\hline
	I & 4 & 5000 & 4 & 1000 \\ 
	II & 5 & 8000 & 5 & 1000 \\
	III & 10 & 20000 & 10 & 2000 \\
	IV & 5 & 8000 & 5 & 1000 \\
	\hline
\end{tabular}
\caption{Convergence parameters for MF-RPMD and MV-RPMD calculations 
of the nuclear TCFs for Models I-IV.  
We report the number of path beads, $P$,
and the number of trajectories, $N_\text{T}$, used for each simulation.}
\label{table:convergence}
\end{table}
%\vspace{-0.5in}
\end{center}
\vspace{-0.2in}

In Fig.~\ref{fig:single_traj10}, we present the instantaneous values of 
electronic state populations and  nuclear center-of-mass coordinate 
for Model III along a representative trajectory. 
In this strong coupling regime, 
we observe the clear separation between the fast timescales on which 
electronic state populations oscillate about the equilibrium value of $0.5$
and the slower timescale associated with nuclear vibrational motion. 
Comparing Fig.~\ref{fig:single_traj} with Fig.~\ref{fig:single_traj10} we 
observe the change in mechanism from a regime where nonadiabatic
transitions between electronic states are coupled to nuclear vibrations to
the mean-field regime where nuclear motion occurs on an average 
electronic potential surface.

\begin{figure}[!htb]
\vspace{0.1in}
   \includegraphics[angle=0,scale=0.3]{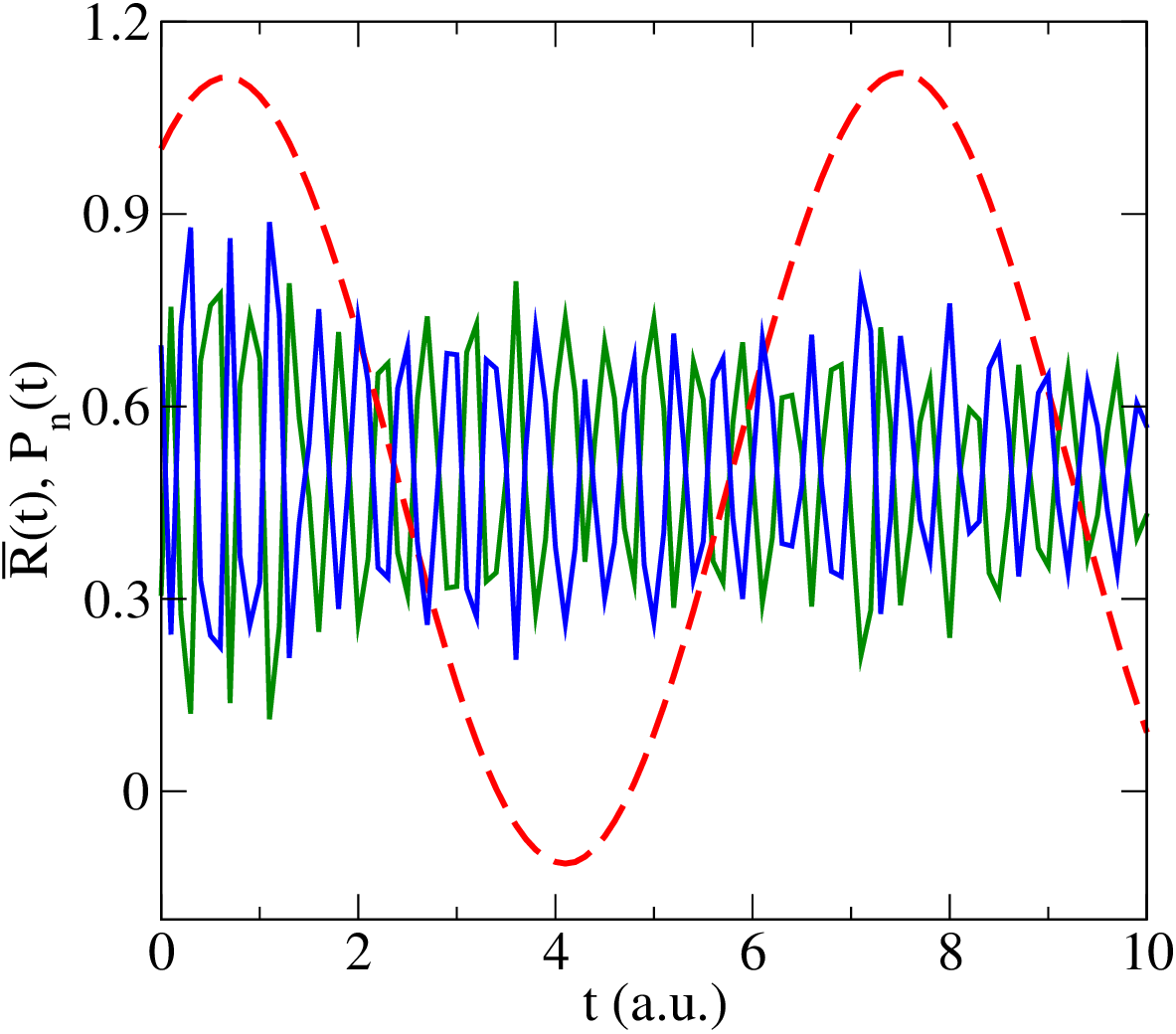}
  \caption{The instantaneous values of the electronic state populations
  for Model III are shown in blue (solid line) for state 1 and 
  in green (solid line) for state 2, and the nuclear center-of-mass
coordinate is shown in red (dashed line) along a single MV-RPMD trajectory.
The nuclear coordinate is scaled and shifted such that the diabatic
  potentials appear to cross at $R=0.5$ rather than $R=0$, 
and we only show the first $10$ a.u. of the trajectory for clarity.}
  \label{fig:single_traj10}
  \vspace{0cm}
\end{figure}

Our asymmetric system, Model IV, has potential energy
diabats as shown in Fig.~\ref{fig1:potential} and is in the weak-coupling 
regime. The system is deliberately chosen to resemble the 
inverted regime in a system-bath model for electron transfer, a case
known to challenge the accuracy of the position-space, nonadiabatic RPMD 
approach\cite{arm11}. In Fig.~\ref{fig:crr_asym3_a1dpt1} we show the 
remarkable agreement between the MV-RPMD TCF and the exact quantum result.
MF-RPMD performs reasonably well, but agrees with the exact results 
only at very short times and fails to capture the correct timescale for 
nuclear motion. 
%The MV-RPMD simulations were converged with $P=5$ 
%beads and $8000$ trajectories.
\begin{figure}[!htb]
   \includegraphics[angle=0,scale=0.3]{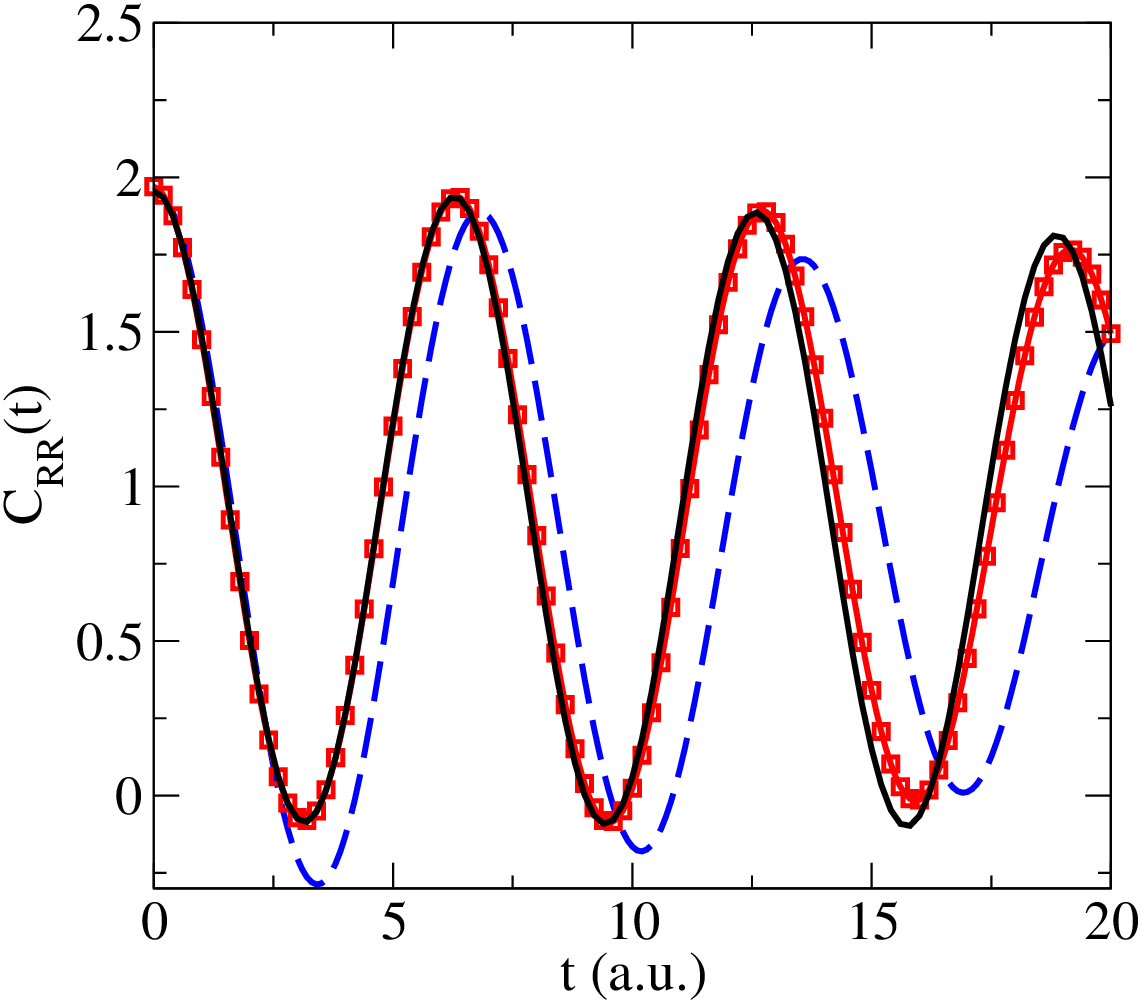}
\caption{The nuclear position-position TCFs for Model IV obtained 
  using MV-RPMD is shown in red (solid line with squares),
  the TCF using MF-RPMD is shown in blue (dashed line),
  and the exact DVR grid calculation is shown in black (solid line).}
  \label{fig:crr_asym3_a1dpt1}
  \vspace{0cm}
\end{figure}

The number of trajectories, $N_\text{T}$, required to converge 
the TCF calculations for all 
four models described here are recorded in Table~\ref{table:convergence}.
We note that although the computational expense is comparable to a 
low-cost linearized semiclassical implementation,\cite{na10} 
the resulting TCFs are expected to be numerically more 
accurate since MV-RPMD avoids the problem of zero-point energy leakage 
by employing trajectories that preserve the QBD. 
Further, we find the sign function, $\text{sgn}(\Theta)$, in the 
Eq.~(\ref{eq:mvr_tcf_is}) does not give rise to a numerical sign problem
in the calculations presented here, which we attribute to 
the non-oscillatory structure of the QBD from which our 
dynamics is derived.
We also report the number of path-beads, $P$, required to converge  
the TCF calculations in Table~\ref{table:convergence}.
The validity of the short-time approximation for the electronic state matrix
elements in Eq.~(\ref{eq:boltz_mat}) is related to the coupling strength,
as evidenced by an increase in the number of path-beads required 
as we go from the weak-coupling to the strong-coupling regime.
The MV-RPMD approach is thus particularly suited for simulating chemistry
in the weak-coupling limit. It is notable that the MV-RPMD method 
is able to accurately describe nonadiabatic dynamics over a wide range of 
coupling strengths and also correctly describes asymmetric tunneling. 

Looking forward, for the calculation of chemical reaction rates in general 
nonadiabatic systems, it will be desirable to identify good order parameters 
in the electronic state variables as well as in nuclear coordinates.
In addition, while the MV-RPMD method is very promising for applications to 
photochemical processes, the fundamental success of RPMD in 
describing quantum dynamics, despite recent progress, \cite{sh13,tjhh13} 
is not fully understoond and requires further theoretical investigation.

\section{Conclusion}
In this paper we derive a novel imaginary-time PI based dynamics, 
MV-RPMD, that extends the applicability of RPMD to multi-electron 
processes. Using standard benchmark models for nonadiabatic systems, 
we demonstrate that the MV-RPMD dynamics are capable of accurately
simulating TCFs for model systems with coupling strengths that range 
over two orders of magnitude.
We also demonstrate that the method is robust to asymmetries in the 
diabatic electronic states. 

We expect this model dynamics to open the door to large-scale simulations 
of photo-induced charge and energy transfer in the condensed phase. Future
applications include exciton dynamics in photovoltaic materials as well
as mechanistic studies of 
multi-electron transfer reactions in transition-metal catalysts.

\section{Acknowledgements}
The author sincerely thanks T. F. Miller III for several valuable 
discussions. This work was supported by a start-up
grant from Cornell University.
\section{Appendix: Wigner transform of the electronic integral}
We start with Eq.~(\ref{eq:psp}) for the electronic integral,
\begin{eqnarray}
\label{eq:psp2}
\nonumber
&&I_E=
\frac{1}{(2\pi\hbar)^{PN}}
\int d\{\xvec_\alpha\}
\int d\{\pvec_\alpha\}
\int d\{\Delta\xvec_\alpha\}\\
&&\;\;
\nonumber
\prod_{\alpha=1}^P
\bra{\xvec_\alpha-\frac{\Delta\xvec_\alpha}{2}}
\proj e^{\beta_P \mathbf{V}(\rvec_\alpha)}\proj
\ket{\xvec_{\alpha+1}+\frac{\Delta\xvec_{\alpha+1}}{2}}
\;e^{i\pvec_\alpha^T\cdot\Delta\xvec_\alpha},
\end{eqnarray}
and substitute Eq.~(\ref{eq:seo_wvfn}) and 
Eq.~(\ref{eq:boltz_mat}) to obtain
\begin{eqnarray}
\nonumber
I_E&=&
\frac{1}{(2\pi\hbar)^{PN}}
\int d\{\xvec_\alpha\}
\int d\{\pvec_\alpha\}
\int d\{\Delta\xvec_\alpha\}\\
\nonumber
&\times&
\prod_{\alpha=1}^P 
\left( \xvec_\alpha-\frac{\Delta\xvec_\alpha}{2} \right)^T\cdot
\intmat(\rvec_\alpha)\cdot
\left( \xvec_{\alpha+1}+\frac{\Delta\xvec_{\alpha+1}}{2} \right)\\
\label{eq:like_pist_prodform}
&\times&
e^{-\sum_{\alpha=1}^P
\left( \frac{1}{4}
\Delta\xvec_\alpha^T\cdot\Delta\xvec_\alpha+
\xvec_\alpha^T\cdot\xvec_\alpha+ 
\pvec_\alpha^T\cdot\pvec_\alpha
-i\pvec_\alpha^T\cdot\Delta\xvec_\alpha\right)}.
\end{eqnarray}
The pre-exponential function in Eq.~(\ref{eq:like_pist_prodform})
can be re-arranged to group like terms in $\{\Delta\xvec_\alpha\}$,
\begin{eqnarray}
\nonumber
I_E&=&
\frac{1}{(2\pi\hbar)^{PN}}
\int d\{\xvec_\alpha\}
\int d\{\pvec_\alpha\}
\int d\{\Delta\xvec_\alpha\}\\
\nonumber
&\times&
\text{Tr}\left[ 
\prod_{\alpha=1}^P
\left( \xvec_\alpha+\frac{\Delta\xvec_\alpha}{2} \right)\otimes
\left( \xvec_\alpha-\frac{\Delta\xvec_\alpha}{2} \right)^T
\intmat(\rvec_\alpha)\right]\\
\label{eq:like_pist_prodform2}
&\times&
e^{-\sum_{\alpha=1}^P-
\left( \frac{1}{4}
\Delta\xvec_\alpha^T\cdot\Delta\xvec_\alpha+
\xvec_\alpha^T\cdot\xvec_\alpha+ 
\pvec_\alpha^T\cdot\pvec_\alpha
-i\pvec_\alpha^T\cdot\Delta\xvec_\alpha\right)}.
\end{eqnarray}
The resulting integral over $\{\Delta\xvec_\alpha\}$
is of Gaussian form and can be analytically evaluated 
to obtain Eq.~(\ref{eq:psint_ie}).

%%fakesection: bibliography

\end{document}